\documentclass[12pt,a4paper]{article}

\usepackage{subfigure}
\usepackage{array}
\usepackage{graphicx}
\usepackage{amsmath}
\usepackage{amssymb}
\usepackage{mathrsfs}
\usepackage{bm}
\usepackage{slashed}
\usepackage{threeparttable}
\usepackage[colorlinks, linkcolor=black, anchorcolor=black, citecolor=black]{hyperref}
\usepackage[amsmath,thmmarks,hyperref]{ntheorem}
\usepackage{soul}
\usepackage{authblk}

\title{Annihilation Rates of $^3D_2(2^{--})$ and $^3D_3(3^{--})$ Heavy Quarkonia}

\author{Tianhong Wang$^a$\footnote{thwang@hit.edu.cn},~~Hui-Feng Fu$^b$,~Yue Jiang$^a$,~Qiang Li$^a$
\\
\vspace{-0.25cm}and~~Guo-Li Wang$^a$\footnote{gl\_wang@hit.edu.cn}\\
{\it \small   $^a$Department of Physics, Harbin Institute of Technology,
Harbin, 150001, China}
\\ \it\small  $^b$Center for Theoretical Physics, College of Physics, Jilin University, Changchun, 130012, China}

\begin{document}
\maketitle

\begin{abstract}
We calculate the annihilation decay rates of the $^3D_2(2^{--})$ and $^3D_3(3^{--})$ charmonia and bottomonia by using the instantaneous Bethe-Salpeter method. The wave functions of states with quantum numbers $J^{PC}=2^{--}$ and $3^{--}$ are constructed. By solving the corresponding instantaneous Bethe-Salpeter equations, we obtain the mass spectra and wave functions of the quarkonia. The annihilation amplitude is written within Mandelstam formalism and the relativistic corrections are taken into account properly. This is important, especially for high excited states, since their relativistic corrections are large. The results for the $3g$ channel are as follows: $\Gamma_{^3D_2(c\bar c)\rightarrow ggg} = 9.24$ keV, $\Gamma_{^3D_3(c\bar c)\rightarrow ggg}=25.0$ keV, $\Gamma_{^3D_2(b\bar b)\rightarrow ggg}= 1.87$ keV, and $\Gamma_{^3D_3(b\bar b)\rightarrow ggg}= 0.815$ keV.
\end{abstract}

\section{Introduction}

The $1{^3D_2}(2^{--})$ charmonium has been found in $B$ decays by the Belle Collaboration~\cite{Belle}. It was confirmed very recently by the BESIII Collaboration through the $e^+e^-$ annihilation process with a statistical significance of $6.2 \sigma$~\cite{BES}. The mass of this particle is measured to be $3821.7\pm1.3\pm 0.7$ MeV, and the decay width is less than $16$ MeV. The discovery of this triplet $D$-wave charmonium is important for checking the validity of phenomenological models, such as the quark potential models, which have predicted abundant heavy quarkonium spectra~\cite{God}.

These experimental results have attracted some theoretical attention to the production properties of this particle, such as the possibility to find this particle through $B_c$ decays~\cite{Sang} or $e^+e^-$ annihilation with soft pion limit~\cite{Volo}.
For the decay properties of this particle, since the mass of this particle is below the $D\bar{D}^\ast$ threshold, and the $D\bar D$ channel is forbidden, there is no OZI-allowed channel. As a result, one-photon radiation processes~\cite{Ei} and decays to light hadrons~\cite{He} are important. The later one is closely related to the three-gluon annihilation process. This channel is expected to have a relatively small partial width as it is in order of $\alpha_s^3$. Nevertheless, it still deserves a careful investigation, as it gives useful information to understand the formalism of quark-antiquark interaction and provides a testing ground for the non-perturbative properties of QCD.

For similar reasons, annihilation processes of $D$-wave quarkonia with $J^{PC}=3^{--}$ also need investigations. The $1{^3D_3}(3^{--})$ charmonium has not been found experimentally, and its mass is predicted to be $3812\sim 3903$ MeV by potential models~\cite{Li1}. Although the $D\bar D$ channel of this particle is opened, the high partial wave contribution makes it suppressed. In the bottomonium sector, only the $1{^3D_2}$ state has been found~\cite{CLEO, PDG}. The mass of $1{^3D_3}$ state is predicted to be $10.181$ GeV by Lattice QCD~\cite{Lattice}, and $10.16$ GeV~\cite{God} by potential models. Both states are below the open-flavor-decay threshold.

The annihilation processes of $^3D_2$ and $^3D_3$ quarkonium states have been investigated only in a few works. Refs.~\cite{Ber, Volk, Bel} employed non-relativistic models to calculate the annihilation amplitudes, which, for $D$-wave states, are only related to the second derivative of the wave functions at the origin. Ref.~\cite{He} used the NRQCD method to calculate annihilation decay widths. Since the relativistic corrections to the three-gluon annihilation processes of quarkonia are large~\cite{Fu, Chao}, especially, the non-original parts of the wave functions give considerable contributions, it is important at this stage to investigate the three-gluon annihilation processes of $2^{--}$ and $3^{--}$ $D$-wave quarkonia with relativistic corrections taken into account. In our previous work~\cite{Fu}, the three-gluon (photon) annihilation process of $^3S_1(1^{--})$ charmonia and bottomonia have been calculated with an instantaneous Bethe-Salpeter (BS) method~\cite{BS1, BS2}, and the obtained decay widths are within the limits of experimental error~\cite{BES01}. So in this work, we use the same framework as the one used in Ref.~\cite{Fu} to calculate annihilations of $^3D_2$ and $^3D_3$ charmonium and bottomonium states, that is, we construct the Salpeter wave functions for these mesons and write the decay amplitude within Mandelstam formalism~\cite{Man}.

The remaining of this paper is organized as follows. In Section 2, we present the details of the theoretical formalism including the wave functions and the decay amplitude. Numerical results and discussions for the the annihilation processes of $^3D_2$ and $^3D_3$ heavy quarkonia are presented in Section 3. Section 4 is devoted to a summary. The eigenvalue equations fulfilled by $^3D_2$ and $^3D_3$ heavy mesons are given in the Appendix.

\section{Theoretical calculations}

The $ggg$ and $\gamma gg$ decay widths of the $^3D_2$ and $^3D_3$ mesons are related to that of the three-photon channel just by a parameter. So here we first calculate the later case. According to the Mandelstam formalism~\cite{Man}, the three-photon annihilation amplitude (see Fig. 1) is written as
\begin{equation}\label{T3}
\begin{aligned}
T_{3\gamma} &= \sqrt{3}(iee_q)^3\int\frac{d^4q}{(2\pi)^4}{\rm Tr}\Bigg[\chi_P(q)\Bigg(\slashed\epsilon_3\frac{1}{\slashed k_3-\slashed p_2-m_q+i\epsilon}\slashed\epsilon_2\frac{1}{\slashed p_1-\slashed k_1-m_q+i\epsilon}\slashed\epsilon_1\\
&+ {\rm all~other~permutations~of}~1, 2, 3 \Bigg)\Bigg],
\end{aligned}
\end{equation}
where $\sqrt{3}$ is the color factor; $ee_q$ is the electric charge of the heavy quark in unit of $e$ (for charmonium $e_q=\frac{2}{3}$ and for bottomonium $e_q=-\frac{1}{3}$); $\chi_P(q)$ is the Bethe-Salpeter wave function of the meson with mass $M$ and momentum $P$, and $q$ is the relative momentum of the inner quark and antiquark (with mass $m_q$ and momentum $p_i$); $k_1\sim k_3$ are momenta of final photons (gluons) with polarizations $\epsilon_1\sim\epsilon_3$, respectively.

\begin{figure}[ht]
\centering
\subfigure[]{\includegraphics[scale=0.8]{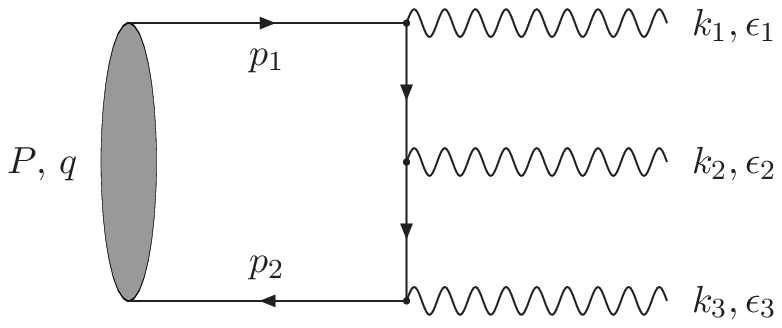}}\\
\subfigure[]{\includegraphics[scale=0.8]{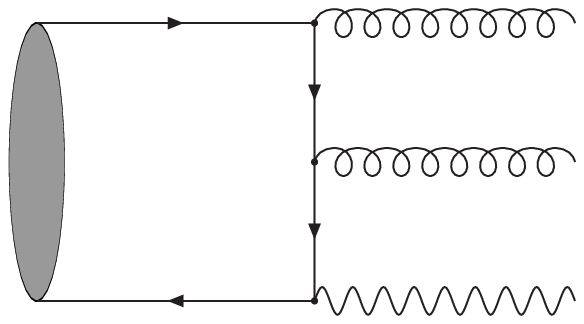}}
\hspace{15mm}
\subfigure[]{\includegraphics[scale=0.8]{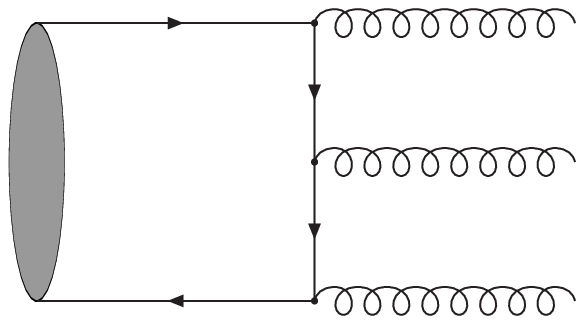}}
\caption[]{Feynman diagrams for the annihilation processes: (a) $\gamma\gamma\gamma$; (b)$\gamma gg$; (c)$ggg$. For each case, there are also five other diagrams with permutations of photons and gluons.}
\end{figure}

To do the integration in Eq.~(\ref{T3}), we take the approximation $p_1\rightarrow \widetilde{p}_1=\frac{1}{2}P+q_\perp$ and $p_2\rightarrow \widetilde{p}_2=\frac{1}{2}P-q_\perp$ ($q_\perp$ is defined as $q-\frac{P\cdot q}{\sqrt{P^2}}P$), which is reasonable when $p_1^0+p_2^0\approx M$. By doing so, the heavy quark propagators will only depend on $\vec q$, while $q^0$ is only included in the wave function. By using the definition
\begin{equation}
\varphi_P(q_\perp)=i\int\frac{dq^0}{2\pi}\chi_P(q),
\end{equation}
we could get the three-dimensional form of the amplitude
\begin{equation}
\begin{aligned}
T_{3\gamma}&=\sqrt{3}(iee_q)^3\int\frac{d\vec q}{(2\pi)^3}{\rm Tr}\Bigg\{\varphi_P(q_\perp)\Bigg[\slashed\epsilon_3\frac{\slashed k_3-\widetilde{\slashed p}_2+m_q}{(k_3-\widetilde{p}_2)^2-m_q^2 +i\epsilon}\slashed\epsilon_2\frac{\widetilde{\slashed p}_1-\slashed k_1 +m_q}{(\widetilde{p}_1-k_1)^2-m_q^2+i\epsilon}\slashed\epsilon_1\\
&+ {\rm all ~other ~permutations ~of} ~1, 2, 3\Bigg]\Bigg\}.
\end{aligned}
\end{equation}

Here we give the explicit expressions for the $D$-wave mesons. Following Refs.~\cite{Kim, Wang}, with the instantaneous approximation (set $q^0=0$), the general  wave function of $^3D_2$ meson is constructed to have the following form
\begin{equation}
\begin{aligned}
\varphi_{2^{--}}(q_\perp) = i\epsilon_{\mu\nu\alpha\beta}\frac{P^\nu}{M}q_\perp^\alpha\epsilon^{\beta\delta}q_{\perp\delta}\gamma^\mu \left(g_1 + \frac{\slashed P}{M} g_2 + \frac{\slashed q_\perp}{M}g_3 + \frac{\slashed P\slashed q_\perp}{M^2} g_4\right),
\end{aligned}
\end{equation}
where $\epsilon^{\mu\nu}$ is the polarization tensor of the meson and $\epsilon_{\mu\nu\alpha\beta}$ is the {\it Levi-Civita} simbol; $g_i$s are functions of $q^2_\perp$, one can check that this wave function has the quantum number of $J^{PC}=2^{--}$. For the $^3D_3$ state, according to the quantum number $3^{--}$, its wave function is given as follows
\begin{equation}
\begin{aligned}
\varphi_{3^{--}}(q_\perp)& = \epsilon_{\mu\nu\alpha}q_\perp^\mu q_\perp^\nu\left[q_\perp^\alpha\left(f_1 + \frac{\slashed P}{M} f_2 + \frac{\slashed q_\perp}{M}f_3 + \frac{\slashed P\slashed q_\perp}{M^2} f_4\right)+M\gamma^\alpha \left(f_5 + \frac{\slashed P}{M} f_6 \right.\right.\\
&\left.\left.+ \frac{\slashed q_\perp}{M}f_7 + \frac{\slashed P\slashed q_\perp}{M^2} f_8\right)\right],
\end{aligned}
\end{equation}
where $\epsilon^{\mu\nu\alpha}$ is the third-order polarization tensor of the meson. As there are constrained conditions (see Appendix), not all the $g_i$s and $f_i$s are independent. For the $2^{--}$ state, only $g_1$ and $g_2$ are independent, while for the $3^{--}$ state, $f_3\sim f_6$ are independent. The numerical values of these independent wave functions can be obtained by solving the Salpeter equations, and the corresponding eigenvalue equations and normalization conditions are given in the Appendix.

The three-photon decay width is given by
\begin{equation}
\begin{aligned}
\Gamma_{3\gamma}=\frac{1}{3!}\frac{1}{8M(2\pi)^3}\int_{0}^{\frac{M}{2}}dk_1\int_{\frac{M}{2}-k_1}^{\frac{M}{2}}dk_2\frac{1}{2J+1}\sum_{{\rm pol}}|T_{3\gamma}|^2,
\end{aligned}
\end{equation}
where $J$ is the spin of the meson. To sum the meson polarization, we have used the complete relation of polarization tensors~\cite{Ber2}. First we difine
\begin{equation}
\mathcal{P}_{\mu\nu} \equiv-g_{\mu\nu} + \frac{P_\mu P_\nu}{M^2}.
\end{equation}
For the $2^{--}$ state, the relation is
\begin{equation}
\begin{aligned}
\sum_{\lambda}\epsilon_{\mu\nu}^{(\lambda)}\epsilon_{\mu^\prime\nu^\prime}^{\ast(\lambda)} &=\frac{1}{2}\left(\mathcal{P}_{\mu\mu^\prime}\mathcal{P}_{\nu\nu^\prime}+\mathcal{P}_{\mu\nu^\prime}\mathcal{P}_{\nu\mu^\prime}\right)-\frac{1}{3}\mathcal{P}_{\mu\nu}\mathcal{P}_{\mu^\prime\nu^\prime},
\end{aligned}
\end{equation}
and for the $3^{--}$ state, it has the form
\begin{equation}
\begin{aligned}
\sum_{\lambda} \epsilon^{(\lambda)}_{abc}\epsilon^{\ast(\lambda)}_{xyz}&=\frac{1}{6}(\mathcal{P}_{ax}\mathcal{P}_{by}\mathcal{P}_{cz}+\mathcal{P}_{ax}\mathcal{P}_{bz}\mathcal{P}_{cy}+\mathcal{P}_{ay}\mathcal{P}_{bx}\mathcal{P}_{cz}\\
&+\mathcal{P}_{ay}\mathcal{P}_{bz}\mathcal{P}_{cx}+\mathcal{P}_{az}\mathcal{P}_{by}\mathcal{P}_{cx}+\mathcal{P}_{az}\mathcal{P}_{bx}\mathcal{P}_{cy})\\
&-\frac{1}{15}(\mathcal{P}_{ab}\mathcal{P}_{cz}\mathcal{P}_{xy}+\mathcal{P}_{ab}\mathcal{P}_{cy}\mathcal{P}_{xz}+\mathcal{P}_{ab}\mathcal{P}_{cx}\mathcal{P}_{yz}\\
&+\mathcal{P}_{ac}\mathcal{P}_{bz}\mathcal{P}_{xy}+\mathcal{P}_{ac}\mathcal{P}_{by}\mathcal{P}_{xz}+\mathcal{P}_{ac}\mathcal{P}_{bx}\mathcal{P}_{yz}\\
&+\mathcal{P}_{bc}\mathcal{P}_{az}\mathcal{P}_{xy}+\mathcal{P}_{bc}\mathcal{P}_{ay}\mathcal{P}_{xz}+\mathcal{P}_{bc}\mathcal{P}_{ax}\mathcal{P}_{yz}).
\end{aligned}
\end{equation}

For the decay channels $^3D_2(^3D_3)\rightarrow \gamma gg$ and $^3D_2(^3D_3)\rightarrow ggg$, the decay widths are~\cite{Fu}
\begin{equation}
\begin{aligned}
\Gamma_{\gamma gg}= \frac{2}{3}\frac{\alpha_s^2}{\alpha^2e_q^4}\Gamma_{3\gamma},
\end{aligned}
\end{equation}
and~\cite{Nov}
\begin{equation}
\begin{aligned}
\Gamma_{ggg}= \frac{5}{54}\frac{\alpha_s^3}{\alpha^3e_q^6}\Gamma_{3\gamma},
\end{aligned}
\end{equation}
respectively.

\section{Results and Discussions}

When solving the Salpeter equation, we use the instantaneously approximated potential which in the momentum space has the following form
\begin{equation}
\begin{aligned}
\label{Cornell}
V(\vec{q})=(2\pi)^3V_s(\vec{q})
+\gamma_0\otimes\gamma^0 (2\pi)^3 V_v(\vec{q}),\\
V_{s}(\vec{q})
=-\left(\frac{\lambda}{\alpha}+V_0\right)\delta^{3}(\vec{q})
+\frac{\lambda}{\pi^{2}}\frac{1}{(\vec{q}^{2}+\alpha^{2})^{2}},\\
V_v(\vec{q})=-\frac{2}{3\pi^{2}}
\frac{\alpha_{s}(\vec{q})}{\vec{q}^{2}+\alpha^{2}},\\
\alpha_s(\vec{q})=\frac{12\pi}{(33-2N_f)}
\frac{1}{{\rm{ln}}\left(a+\frac{\vec{q}^2}{\Lambda^2_{_{QCD}}}\right)}.
\end{aligned}
\end{equation}
Parameters in the above equations have the values~\cite{Fu, Wang}: $a=e=2.71828$, $\alpha=0.06$ GeV, $\lambda=0.21$ ${\rm GeV}^2$, $\Lambda_{QCD}=0.27$ GeV ($0.20$ GeV for $b\bar b$), $m_b=4.96$ GeV, $m_c=1.62$ GeV. We set the flavor number $N_f=3$ for charmonia and $N_f=4$ for bottomonia. By using the fourth equation above we get $\alpha_s(m_c)=0.38$ and $\alpha_s(m_b)=0.23$. We choose appropriate values of $V_0$ to get the mass spectra and wave functions. The results are as follows: $M_{1{^3D_2}(c\bar c)}=3.8217$ GeV, $M_{1{^3D_3}(c\bar c)}=3.830$ GeV, $M_{1{^3D_2}(b\bar b)}=10.1637$ GeV and $M_{1{^3D_3}(b\bar b)}=10.165$ GeV. We want to stress that the potential used here just a phenomenological one. The first term which is a linear potential corresponds to the non-perturbative effects. The second term, which is a Coulomb-like potential, comes from the one-gluon exchange interaction. Here we just keep the time-like part which gives the largest contribution. For the space-like parts, which correspond to the relativistic corrections, are neglected at this stage (to solve the equation with these terms will be done in our future work). So here we can only study the relativistic effect comes from the wave-function with a specific potential form.

The wave functions are plotted in Fig.~2. For the $2^{--}$ state, $g_1$ and $g_2$ are independent functions. The numerical result shows that this two functions are very close to each other. So here we just plot $g_1$ as an example. For the $3^{--}$ state, there are four independent functions, while $f_3$ is close to $f_4$ and $f_5$ is close to $f_6$. So we only plot the $f_3$ and $f_5$. To make the wave functions to be dimensionless, we have rescaled them by a factor. Because the  normalization condition is different, this factor for $2^{--}$ and $3^{--}$ is different. One can see the wave functions of $b\bar b$ are quite large than those of $c\bar c$. This is mainly because we have used different scale factors. Actually, for $2^{--}$ states, $g_i (b\bar b)$ is more than two times smaller than $g_i (c\bar c)$. It should be mentioned that the position of the peak value for the former is at the right of the later. This means that the contribution coming from non-zero $|\vec q|$ for the $b\bar b$ state is larger than that for the $c\bar c$ state.

\begin{figure}[ht]
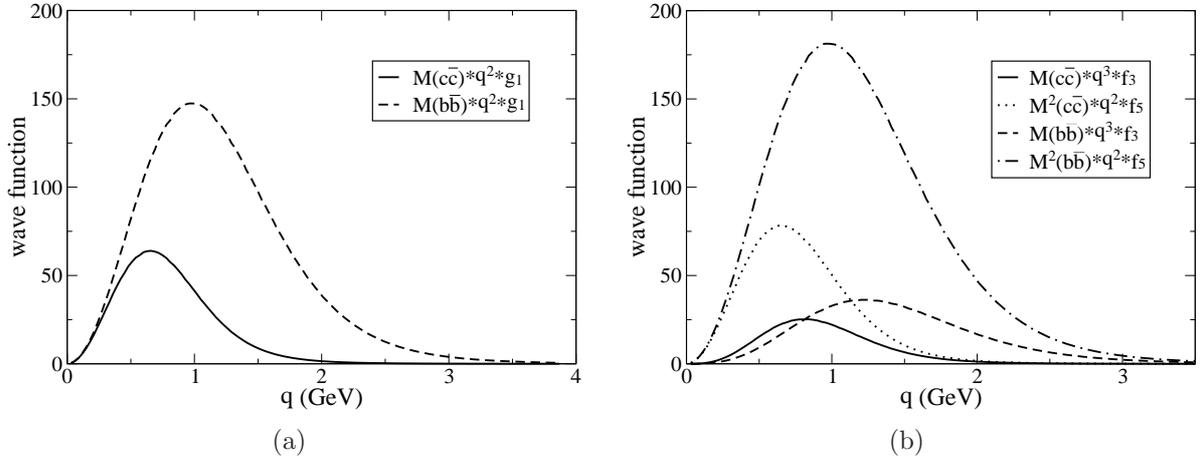
\label{wf}
\centering
\subfigure[]{\includegraphics[scale=0.31]{2--.eps}}
\hspace{3mm}
\subfigure[]{\includegraphics[scale=0.31]{3--.eps}}
\caption[]{Wave functions of $2^{--}$ and $3^{--}$ mesons. (a) for the $2^{--}$ state, only $f_1$ is plotted both for charmonium and bottomonium. (b) for the $3^{--}$ states, the $f_3$ and $f_5$ are presented. All the wave functions are rescaled to be dimensionless.}
\end{figure}

In Fig.~3, we present the three-photon differential decay widths of $2^{--}$ and $3^{--}$ charmonia against $k_1$ and $k_2$ (here we use $k_i$ to represent the photon energy; the projection to the $k_1$-$k_2$ plane is the Dalitz plot). For bottomonia, the results are plotted in Fig.~4. One notices that for the $3^{--}$ states, as the kinematics $(k_1, k_2)$ goes toward the points $(0, \frac{M}{2})$, $(\frac{M}{2}, 0)$, and $(\frac{M}{2}, \frac{M}{2})$, respectively, the differential width gets larger and larger, while for the $2^{--}$ states, only points near $(\frac{M}{2}, \frac{M}{2})$ increases obviously. In Ref.~\cite{Ber}, a similar diagram was given, while there $k_2$ was integrated out. The differential decay width of the $2^{--}$ $c\bar c$ ($b\bar b$) state is generally smaller than that of the $3^{--}$ $c\bar c$ ($b\bar b$) state at the same kinematic point. Compared to the charmonium states, the differential widths of the bottomonia are more flat in the central region.

The decay widths for the chamonia are presented in Table 1. One can see that the three-photon results are tiny which is hard to be detected in the future. For the $^3D_2\rightarrow ggg$ channel, our result is $9.24$ keV, which is about $2$ times smaller than that of Ref.~\cite{Volk} but close to that of Ref.~\cite{Ber}.  For the $^3D_3\rightarrow ggg$ channel, we get $25.0$ keV, which is about $5$ times smaller than that of Ref.~\cite{Volk} and $2.5$ times smaller than that of Ref.~\cite{Ber}. Ref.~\cite{He} gave the result which is almost $5$ times larger than ours for the $^3D_2\rightarrow ggg$ channel and 9 times for the $^3D_3\rightarrow ggg$ channel. The results of Ref.~\cite{He} we cited here is calculated at $\mu=m_c$. The authors there also gave the widths at $\mu=2m_c$, which are 50 keV and 172 keV for $^3D_2$ and $^3D_3$, respectively.

The decay widths of $^3D_2$ and $^3D_3$ bottomonia are listed in Table 2. For the three-photon decay channels, they are smaller than that of the charmonia by $2\sim 3$ orders of magnitude, while for the other two channels, they are about two orders and one order of magnitude smaller, respectively. {The main reason for this is that the electric charges of heavy quarks differs by a factor of 2, which increases to 64 at the decay width level. Together with the mass difference of the initial mesons, this factor almost get to 200.}  For the channels with gluons, the strong coupling constant $\alpha_s$ has different value at different energy scale, which also causes depression for $b\bar b$. {Our results for $\Gamma_{^3D_2\rightarrow ggg}$ of the bottomonium is roughly 7 (4, 3) times larger than that of Ref.~\cite{Bel} (\cite{Ber}, \cite{He}), while for the $\Gamma_{^3D_3\rightarrow ggg}$ channel, it is 1 (3, 3) times smaller. In Ref.~[14], $\alpha_s$ is taken to be $0.17$ for $\Upsilon(1{\rm D})$, which is smaller than 0.23 in our work (also in Ref.~[12]). As the three-gluon decay width is proportional to $\alpha_s^3$ (see Eq.~(11)), our result will be 3.3 times smaller than that in Ref.~[14] if we take the same value of $\alpha_s$ as that in Ref.~[14]. }

{ To show how large of the relativistic effects, we use the wave functions without relativistic parts (for $2^{--}$, this mainly comes from $g_3$ and $g_4$ parts; for $3^{--}$, this mainly comes form the terms except $f_5$ and $f_6$) to do the similar calculation. The results are presented within parentheses in Table 1 and Table 2. One can see the results with relativistic corrections are smaller than those without relativistic corrections. For the charmonia, these corrections bring $17\%$ and $20\%$ contributions (compared to the data outside parentheses) for $2^{--}$ and $3^{--}$ state, respectively, while for for the bottomonia, the contributions are $10\%$ and $5.7\%$. This means the relativistic corrections in the charmonium decays are larger than those for the bottomonium cases.}

Our result for the ratio of the $\gamma gg$ channel and the $ggg$ channel is
\begin{equation}
\begin{aligned}
\frac{\Gamma(^3D_J\rightarrow \gamma gg)}{\Gamma(^3D_J\rightarrow ggg)} = 6.2\%  ~~~(J=2, 3)
\end{aligned}
\end{equation}
for the charmonium, which is close to $7\%$ given in Ref.~\cite{Ber}. This ratio is totally determined by some basic parameters (the fine structure constant, the strong coupling at the relevant scale, etc.), so it is irrelevant to the model employed and could be used to measure the strong coupling at the corresponding scale. For the bottomonium, our result of this ratio is $2.5\%$ which is also close to $3\%$ in Ref.~\cite{Ber}. As for the ratios of $3g$ channel for different $D$-waves, we get
\begin{equation}
\begin{aligned}
\frac{\Gamma(^3D_3\rightarrow ggg)}{\Gamma(^3D_2\rightarrow ggg)} ={  2.7 ~~({\rm for} ~c\bar c)~{\rm and}~0.44~({\rm for}~b\bar b).}
\end{aligned}
\end{equation}
The ratio is irrelevant to the strong coupling and only reflects the difference in wave functions between $2^{--}$ and $3^{--}$ states. { Our results of this ratio are smaller than} those of other models ($5\sim 6$ for $c\bar c$ and $4\sim 5$ for $b\bar b$), which indicates that the relativistic corrections to the $2^{--}$ and $3^{--}$ states are different. { The large difference between the ratios of $c\bar c$ and $b\bar b$ indicates that the decay amplitudes of $2^{--}$ and $3^{--}$ states change differently when the heavy quark flavor is changed. }

In conclusion, we have calculated the three-photon (gluon) decay widths with the Bethe-Salpeter method {with which the relativistic corrections are taken into account properly}. Our results show that three photon decay channels have very small decay widths, especially for the bottomonium state.  For the three-gluon processes we get: $\Gamma_{3g}[^3D_2,~^3D_3]= { (9.24,~25.0)}$ keV for the charmonia and ${ (1.87,~0.815)}$ keV for the bottomonia. { Compared to the results given by the non-relativistic models, our results are small for the $2^{--}$($c\bar c$) and $3^{--}$($c\bar c$, $b\bar b$) cases, while for the $2^{--}$($b\bar b$) meson, our results are larger than those of other models. Our results also indicate that the three-gluon (photon) annihilation processes of heavy quarkonia suffer large relativistic corrections for the $^3D_2$ and $^3D_3$ states, especially for the charmonium case.}

\begin{figure}[ht]
\centering
\subfigure[]{\includegraphics[scale=0.41]{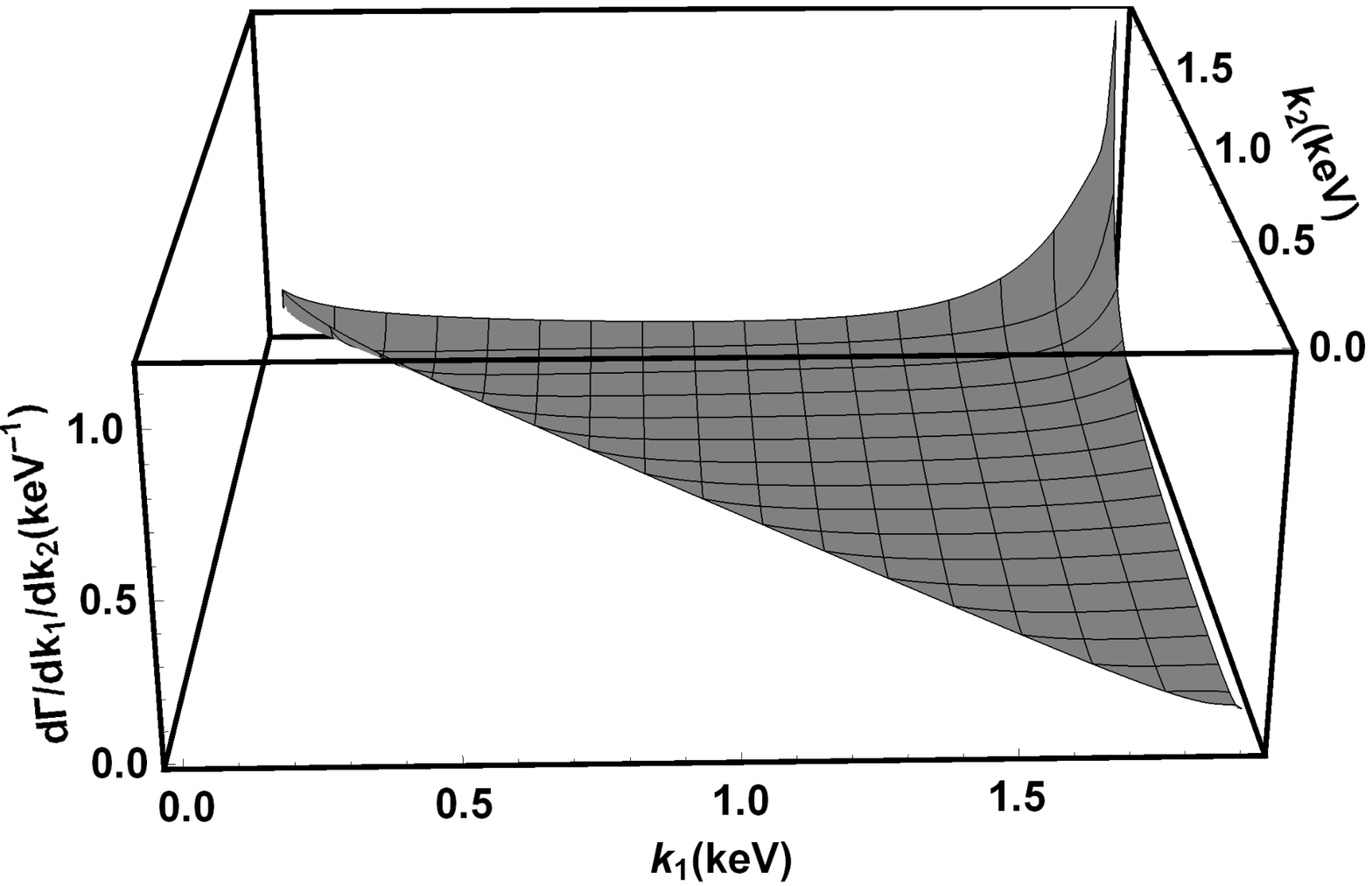}}
\hspace{3mm}
\subfigure[]{\includegraphics[scale=0.41]{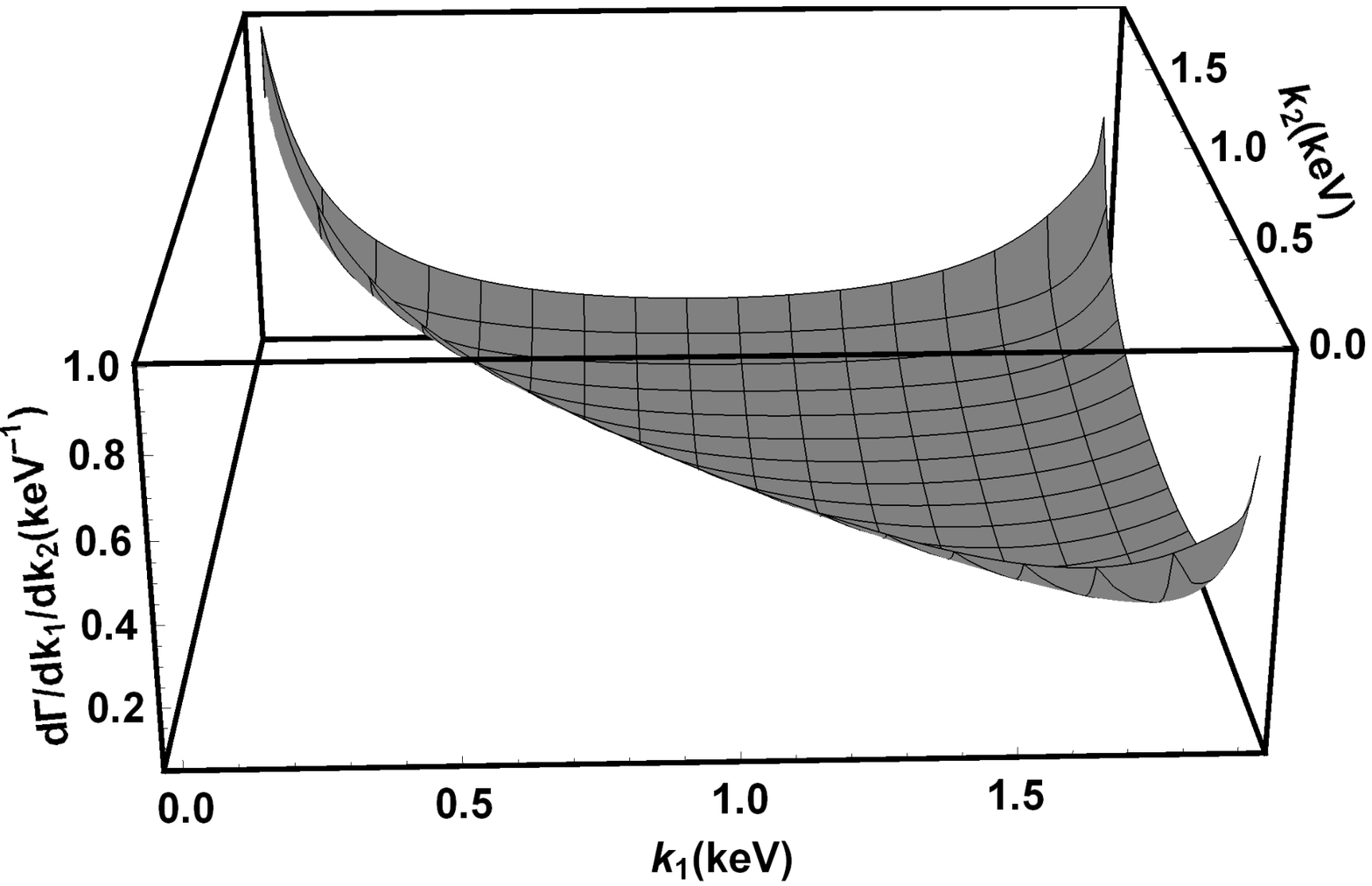}}
\caption[]{The differential width $\frac{d\Gamma}{dk_1dk_2}$ of three-photon decay changes with respect to $k_1$ and $k_2$. (a) for $1{^3D_2}(c\bar c)$ and (b) for $1{^3D_3}(c\bar c)$.}
\end{figure}

\begin{figure}[ht]
\centering
\subfigure[]{\includegraphics[scale=0.41]{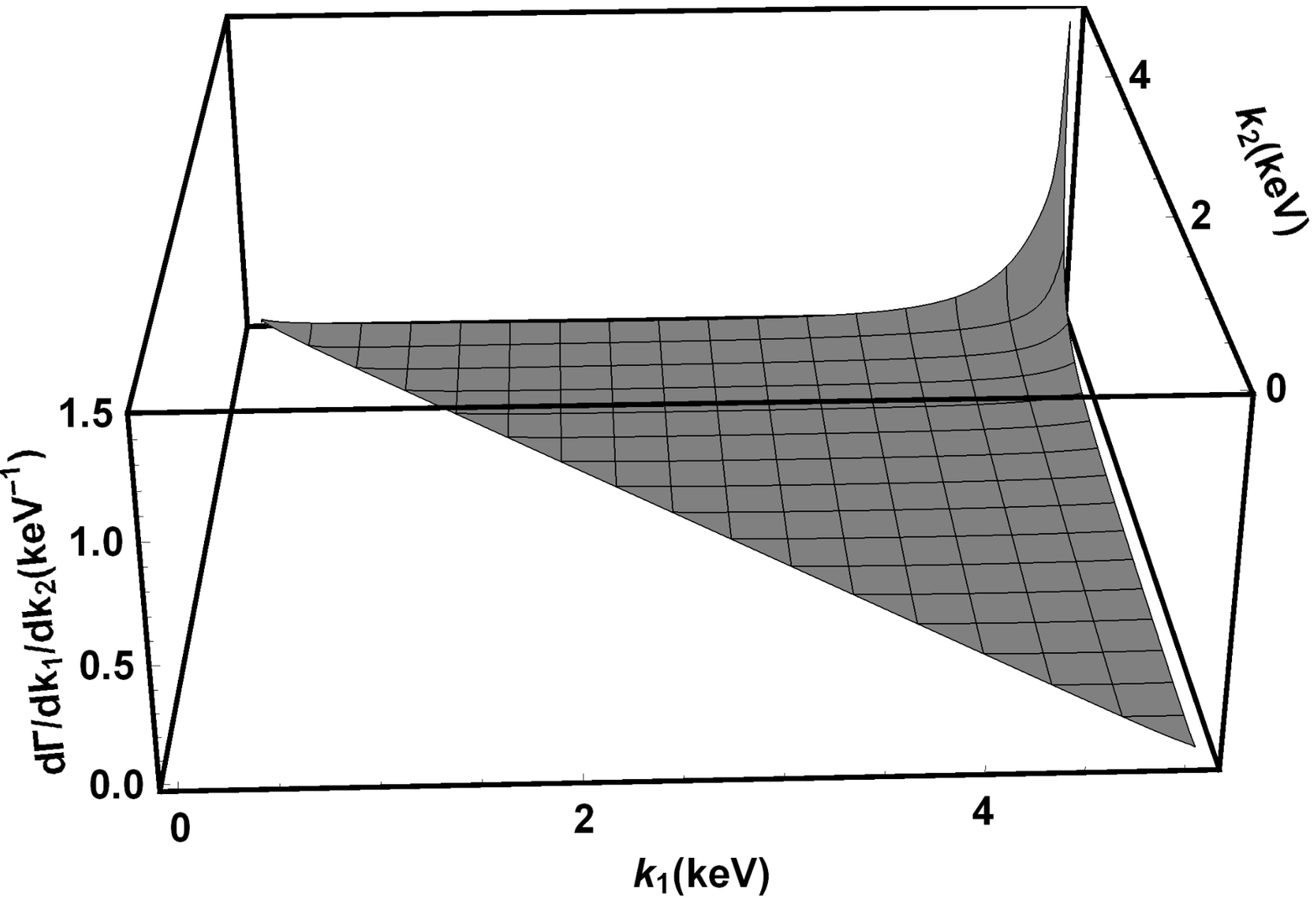}}
\hspace{3mm}
\subfigure[]{\includegraphics[scale=0.41]{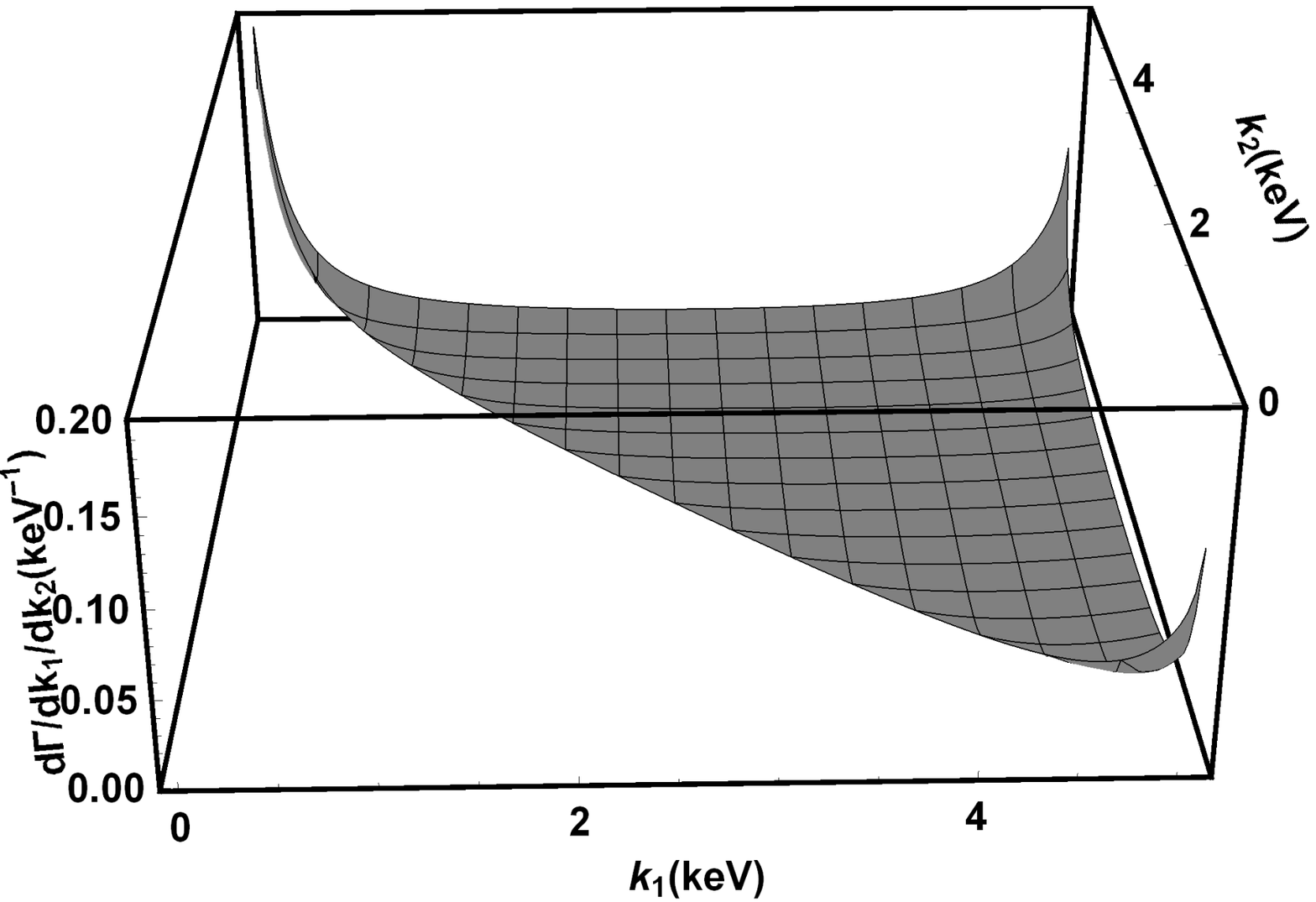}}
\caption[]{The differential width $\frac{d\Gamma}{dk_1dk_2}$ of three-photon decay changes with respect to $k_1$ and $k_2$. (a) for $1{^3D_2}(b\bar b)$ and (b) for $1{^3D_3}(b\bar b)$.}
\end{figure}

\begin{table}[ht]
 \caption{Partial decay widths (keV) of $^3D_2$ and $^3D_3$ charmonia.}
 \setlength{\tabcolsep}{0.1cm}
 \centering
\begin{tabular*}{\textwidth}{@{}@{\extracolsep{\fill}}ccccc}
\hline \hline
Decay Channel & Ours& Ref.~\cite{Volk}&Ref.~\cite{Ber}&Ref.~\cite{He}  \\ \hline
{\phantom{\Large{l}}}\raisebox{+.2cm}{\phantom{\Large{j}}}
$\Gamma_{^3D_2\rightarrow \gamma\gamma\gamma}$ & $6.20(7.25)\times 10^{-5}$&& &\\
{\phantom{\Large{l}}}\raisebox{+.2cm}{\phantom{\Large{j}}}
$\Gamma_{^3D_3\rightarrow \gamma\gamma\gamma}$&$1.68(2.05)\times 10^{-4}$ &&&  \\
{\phantom{\Large{l}}}\raisebox{+.2cm}{\phantom{\Large{j}}}
$\Gamma_{^3D_2\rightarrow \gamma gg}$ & $0.568(0.664)$& &0.84& \\
 {\phantom{\Large{l}}}\raisebox{+.2cm}{\phantom{\Large{j}}}
 $\Gamma_{^3D_3\rightarrow \gamma gg}$&$1.54(1.88)$ & &4.76& \\
 {\phantom{\Large{l}}}\raisebox{+.2cm}{\phantom{\Large{j}}}
$\Gamma_{^3D_2\rightarrow ggg}$ &$9.24(10.8)$ &$19\pm3$ &$12$& 42\\
 {\phantom{\Large{l}}}\raisebox{+.2cm}{\phantom{\Large{j}}}
$\Gamma_{^3D_3\rightarrow ggg}$ &$25.0(30.5)$ &$121$ &$68$&223 \\
\hline\hline
\end{tabular*}
\end{table}

\begin{table}[ht]
 \caption{Partial decay widths (keV) of $^3D_2$ and $^3D_3$ bottomonia.}
 \setlength{\tabcolsep}{0.1cm}
 \centering
\begin{tabular*}{\textwidth}{@{}@{\extracolsep{\fill}}ccccc}
\hline \hline
Decay Channel & Ours& Ref.~\cite{Bel} &Ref.~\cite{Ber} &Ref.~\cite{He} \\ \hline
{\phantom{\Large{l}}}\raisebox{+.2cm}{\phantom{\Large{j}}}
$\Gamma_{^3D_2\rightarrow \gamma\gamma\gamma}$ &$88.5(97.6)\times 10^{-8}$ && &\\
{\phantom{\Large{l}}}\raisebox{+.2cm}{\phantom{\Large{j}}}
$\Gamma_{^3D_3\rightarrow \gamma\gamma\gamma}$&$3.84(4.06)\times 10^{-7}$ && & \\
{\phantom{\Large{l}}}\raisebox{+.2cm}{\phantom{\Large{j}}}
$\Gamma_{^3D_2\rightarrow \gamma gg}$ &$47.5(52.4)\times 10^{-3}$ & &$1.53\times 10^{-2}$& \\
 {\phantom{\Large{l}}}\raisebox{+.2cm}{\phantom{\Large{j}}}
 $\Gamma_{^3D_3\rightarrow \gamma gg}$&$2.06(2.18)\times 10^{-2}$ & &$8.1\times 10^{-2}$& \\
 {\phantom{\Large{l}}}\raisebox{+.2cm}{\phantom{\Large{j}}}
$\Gamma_{^3D_2\rightarrow ggg}$ &$1.87(2.06)$ &$0.26$ &$0.51$ &0.60\\
 {\phantom{\Large{l}}}\raisebox{+.2cm}{\phantom{\Large{j}}}
$\Gamma_{^3D_3\rightarrow ggg}$ &$0.815(0.862)$ & $1.1$ &$2.7$ &2.85\\
\hline\hline
\end{tabular*}
\end{table}

\section{Acknowledgments}
This work was supported in part by the National Natural Science
Foundation of China (NSFC) under Grant No.~11405037, No.~11575048 and No.~11505039, and in part by PIRS of HIT No.T201405, No.A201409, and No.B201506.

\section{Appendix}
The Bethe-Salpeter equation which describes two-body bound state relativistically has the following form \cite{BS1}
\begin{equation}
\begin{aligned}
S^{-1}_1(p_1)\chi_{_P}(q)S^{-1}_2(-p_2)=i\int\frac{d^4k}{(2\pi)^4} V(P; q,k)\chi_{_P}(k),
\end{aligned}
\end{equation}
where $p_1=\frac{1}{2}P+q_\perp$ and $p_2=\frac{1}{2}P-q_\perp$ are respectively the momenta of quark and antiquark in the bound state; $\chi_{_P}(q)$ is the BS wave function of the bound state; $V(P; q,k)$ is the interaction potential between quark and antiquark. The fermion propagator $S_i(Jp_i)$ ($J=(-1)^{i+1}$, $i=1$ for quark and $i=2$ for antiquark) is defined as
\begin{equation}
\begin{aligned}
&-iJS_i(Jp_i)=\frac{\Lambda^+_i}{p_{i}-\omega_i+i\epsilon}+\frac{\Lambda_i^-}{p_{i}+\omega_i-i\epsilon},
\end{aligned}
\end{equation}
where we have used the projector $\Lambda_i^{\pm}(p^\mu_{i\perp})=\frac{1}{2\omega_i}[\frac{\slashed P}{M}\omega_i\pm(\slashed p_{i\perp} + J m_i)]$. $m_i$ is the (anti)quark mass and $\omega_i$ has the form $\sqrt{m_i^2-q_\perp^2}$. With instantaneous approximation, $V(P; q, k) \approx V(q_\perp, k_\perp)$, we write the integral in Eq.~(15) as  $\eta_{_P}(q_\perp)=\int\frac{d^3\vec k}{(2\pi)^3} V(q_\perp,k_\perp)\varphi_{_P}(k_\perp)$. By introducing the notation $\varphi^{\pm\pm}\equiv\Lambda^\pm_1\frac{\slashed P}{M}\varphi(q_\perp)\frac{\slashed P}{M}\Lambda^\pm_2$, we get the instantaneous form of BS equation, which is also called full Salpeter equation \cite{BS2}
\begin{subequations}
\begin{equation}
\begin{aligned}
(M-\omega_1-\omega_2)\varphi^{++}_{P}(q_\perp)= \Lambda^+_1(q_\perp)\eta_{_P}(q_\perp)\Lambda_2^+(q_\perp),
\end{aligned}
\end{equation}
\vspace{-0.6cm}
\begin{equation}
\begin{aligned}
(M+\omega_1+\omega_2)\varphi^{--}_{P}(q_\perp)= -\Lambda^-_1(q_\perp)\eta_{_P}(q_\perp)\Lambda_2^-(q_\perp),
\end{aligned}
\end{equation}
\vspace{-0.5cm}
\begin{equation}
\begin{aligned}
\varphi_{P}^{+-}(q_\perp) = \varphi_P^{-+}(q_\perp) =0.
\end{aligned}
\end{equation}
\end{subequations}
Here Eq.~(17c) are the constrained conditions, which result in relations between $g_{i}s$ or $f_{i}s$ (see below).
The normalization condition for Salpeter wave functions is~\cite{BS2}
\begin{equation}
\int\frac{d\vec q}{(2\pi)^3}{\rm Tr}\Bigg[\overline\varphi^{++}\frac{\slashed P}{M}\varphi^{++}\frac{\slashed P}{M}- \overline\varphi^{--}\frac{\slashed P}{M}\varphi^{--}\frac{\slashed P}{M}\Bigg]= 2P^0,
\end{equation}
where $\overline\varphi_P(q_\perp)$ is defined as $\gamma^0\varphi_P^\dagger(q_\perp)\gamma^0$.

Eq.~(17b) describes the negative energy part of the wave function which gives small contributions. So it is neglected by many authors in literatures. { In our previous work~\cite{Zhihui}, their contribution in the decays of $B_c$ to $P$-wave charmonium were considered. There those parts contributes roughly $10^{-3}\sim 10^{-2}$ less than that of the positive energy part of the wave functions. We also calculated the semi-leptonic process $\eta_c\rightarrow D_sl\nu_l$ which was not published. There we found the contribution of negative energy part of the wave function was also about $1\%$.  So in some cases these parts indeed could be neglected. But in this work we want to estimate how large of the contribution, so the full Salpeter equations are solved.} By inserting Eq.~(4) into Eq.~(17c), we get the constrained conditions
\begin{equation}
\begin{aligned}
g_3 = \frac{M(\omega_1 - \omega_2)}{m_1\omega_2+m_2\omega_1}g_1,~~~g_4 = \frac{M(\omega_1 + \omega_2)}{m_1\omega_2+m_2\omega_1}g_2.
\end{aligned}
\end{equation}
From Eq.~(17a) and Eq.~(17b) we get the eigenvalue equations fulfilled by $2^{--}$ states
\begin{subequations}
\begin{equation}
\begin{aligned}
\left(M-\omega_1-\omega_2\right) \left(g_1 - \frac{\omega_1+\omega_2}{m_1+m_2}g_2\right) &= \int \frac{d\vec k}{(2\pi)^3}\frac{1}{4\omega_1\omega_2\vec q^4}\left[A_1\left(g_1-\frac{m_1+m_2}{\omega_1+\omega_2}g_2\right) \right.\\
&\left.+A_2\left(g_1-\frac{\omega_1+\omega_2}{m_1+m_2}g_2\right) \right],\\
\end{aligned}
\end{equation}
\vspace{-0.3cm}
\begin{equation}
\begin{aligned}
\left(M+\omega_1+\omega_2\right) \left(g_1 + \frac{\omega_1+\omega_2}{m_1+m_2}g_2\right)& =  \int \frac{d\vec k}{(2\pi)^3}\frac{-1}{4\omega_1\omega_2\vec q^4}\left[A_1\left(g_1+\frac{m_1+m_2}{\omega_1+\omega_2}g_2\right) \right.
\\
&\left.+A_2\left(g_1+\frac{\omega_1+\omega_2}{m_1+m_2}g_2\right) \right],
\end{aligned}
\end{equation}
\end{subequations}
where $g_i$s on the left side of the equation are functions of $-{q^2_\perp}$, while those on the right side are functions of $-{k^2_\perp}$ ($k_\perp\equiv k-\frac{P\cdot k}{\sqrt{P^2}}P$). $A_i$s are defined as
\begin{equation}
\begin{aligned}
A_1&=(m_1m_2+\vec q^2+\omega_1\omega_2)\left[\vec k^2\vec q^2 - 3(\vec k\cdot\vec q)^2\right]\left(V_s-V_v\right),\\
A_2&=\frac{(E_1-E_2)(m_1-m_2)}{m_1E_2+m_2E_1}2(\vec k\cdot\vec q)^3 (V_s+V_v),
\end{aligned}
\end{equation}
where we have used the definition $E_i=\sqrt{m_i^2-k_\perp^2}$. By solving the eigenvalue equation (Eq.~(20)) numerically, we obtain the eigenvalue $M$ and wave functions $g_i s$ with
the normalization condition (Eq.(18))
\begin{equation}
\int\frac{d\vec q}{(2\pi)^3}\frac{8\omega_1\omega_2\vec q^4}{m_1\omega_2+m_2\omega_1}g_1g_2 = -5M.
\end{equation}

For the $3^{--}$ state, the constrained conditions are
\begin{equation}
\begin{aligned}
&f_1= \frac{-\vec q^2f_3(\omega_1+\omega_2) + 2M^2f_5\omega_2}{M(m_1\omega_2+m_2\omega_1)},~~~~f_2=\frac{-\vec q^2f_4(\omega_1-\omega_2)+2M^2f_6\omega_2}{M(m_1\omega_2+m_2\omega_1)}\\
&f_7=\frac{M(\omega_1-\omega_2)}{m_1\omega_2+m_2\omega_1}f_5,~~~~f_8=\frac{M(\omega_1+\omega_2)}{m_1\omega_2+m_2\omega_1}f_6.
\end{aligned}
\end{equation}
And the eigenvalue equations are
\begin{subequations}
\begin{equation}
\begin{aligned}
&\left(M-\omega_1-\omega_2\right)\left[-\frac{\vec q^2}{M^2}\left(f_3+\frac{m_1+m_2}{\omega_1+\omega_2}f_4\right)+\left(f_5-\frac{m_1+m_2}{\omega_1+\omega_2}f_6\right)\right]=\\
&\int \frac{d\vec k}{(2\pi)^3}\frac{1}{4\omega_1\omega_2\vec q^4}\left\{
B_1\left[f_3+\frac{(E_1-E_2)(\omega_1-\omega_2)}{(E_1+E_2)(m_1+m_2)}f_4\right] + B_2\left(f_3+\frac{\omega_1+\omega_2}{m_1+m_2}f_4\right)\right.\\
&\left.+B_3\left[f_5-\frac{(E_1-E_2)(\omega_1-\omega_2)}{(E_1+E_2)(m_1+m_2)}f_6\right]
+B_4\left(f_5 - \frac{\omega_1+\omega_2}{m_1+m_2}f_6\right)\right\},
\end{aligned}
\end{equation}
\vspace{-0.3cm}
\begin{equation}
\begin{aligned}
&\left(M+\omega_1+\omega_2\right)\left[-\frac{\vec q^2}{M^2}\left(f_3-\frac{m_1+m_2}{\omega_1+\omega_2}f_4\right)+\left(f_5+\frac{m_1+m_2}{\omega_1+\omega_2}f_6\right)\right]=\\
&\int \frac{d\vec k}{(2\pi)^3}\frac{-1}{4\omega_1\omega_2\vec q^4}\left\{
B_1\left[f_3-\frac{(E_1-E_2)(\omega_1-\omega_2)}{(E_1+E_2)(m_1+m_2)}f_4\right] + B_2\left(f_3-\frac{\omega_1+\omega_2}{m_1+m_2}f_4\right)\right.\\
&\left.+B_3\left[f_5 + \frac{(E_1-E_2)(\omega_1-\omega_2)}{(E_1+E_2)(m_1+m_2)}f_6\right]
+B_4\left(f_5 + \frac{\omega_1+\omega_2}{m_1+m_2}f_6\right)\right\},
\end{aligned}
\end{equation}
\vspace{-0.3cm}
\begin{equation}
\begin{aligned}
&\left(M-\omega_1-\omega_2\right) \left(f_5-\frac{\omega_1+\omega_2}{m_1+m_2}f_6\right)= \int \frac{d\vec k}{(2\pi)^3}\frac{1}{4\omega_1\omega_2\vec q^4}\left\{C_1\left(f_3+\frac{m_1+m_2}{\omega_1+\omega_2}f_4\right)\right.\\
&\left.+C_2\left[f_5-\frac{(\omega_1+\omega_2)(E_1+E_2)}{(m_1-m_2)(E_1-E_2)}f_6\right]+C_3\left(f_5-\frac{m_1+m_2}{\omega_1+\omega_2}f_6\right)\right\},
\end{aligned}
\end{equation}
\vspace{-0.3cm}
\begin{equation}
\begin{aligned}
&\left(M+\omega_1+\omega_2\right) \left(f_5+\frac{\omega_1+\omega_2}{m_1+m_2}f_6\right)= \int \frac{d\vec k}{(2\pi)^3}\frac{-1}{4\omega_1\omega_2\vec q^4}\left\{C_1\left(f_3-\frac{m_1+m_2}{\omega_1+\omega_2}f_4\right)\right.\\
&\left.+C_2\left[f_5 + \frac{(\omega_1+\omega_2)(E_1+E_2)}{(m_1-m_2)(E_1-E_2)}f_6\right]+C_3\left(f_5 + \frac{m_1+m_2}{\omega_1+\omega_2}f_6\right)\right\},
\end{aligned}
\end{equation}
\end{subequations}
where we have defined
\begin{equation}
\begin{aligned}
&B_1=\frac{(E_1+E_2)(m_1+m_2)}{M^2(m_1E_2+m_2E_1)}\vec k\cdot\vec q\vec k^2\left[3\vec q^2\vec k^2-5\left(\vec k\cdot \vec q\right)^2\right](V_s+V_v),\\
&B_2=-\frac{1}{M^2\vec q^2}(m_1\omega_2+m_2\omega_1)(\vec k\cdot\vec q)^2\left[3\vec q^2\vec k^2 - 5\left(\vec k\cdot\vec q\right)^2\right]\frac{m_1+m_2}{\omega_1+\omega_2}(V_s-V_v),\\
&B_3=-\frac{(E_1+E_2)(m_1+m_2)}{m_1E_2+m_2E_1}\vec k\cdot\vec q\left[3\vec q^2\vec k^2-5\left(\vec k\cdot \vec q\right)^2\right](V_s+V_v),\\
&B_4=(m_1\omega_2+m_2\omega_1)\left[\vec q^2\vec k^2 - 3\left(\vec k\cdot\vec q\right)^2\right]\frac{m_1+m_2}{\omega_1+\omega_2}(V_s-V_v),\\
\end{aligned}
\end{equation}
and
\begin{equation}
\begin{aligned}
&C_1=-\frac{3}{4M^2\vec q^2}\left(\omega_1\omega_2+m_1m_2+\vec q^2\right)\left[\vec k^4\vec q^4-6\vec k^2\vec q^2\left(\vec k\cdot \vec q\right)^2+5\left(\vec k\cdot \vec q\right)^4\right](V_s-V_v),\\
&C_2=-\frac{(m_1-m_2)(E_1-E_2)}{m_1E_2+m_2E_1}\vec k\cdot\vec q\left[3\vec k^2\vec q^2-5\left(\vec k\cdot\vec q\right)^2\right](V_s-V_v),\\
&C_3=\left(\omega_1\omega_2+m_1m_2+\vec q^2\right)\left[\vec k^2\vec q^2 - 3\left(\vec k\cdot \vec q\right)^2\right](V_s-V_v).
\end{aligned}
\end{equation}
In Eq.~(24), $f_i$s on the left side and right side are functions of $-q_\perp^2$ and $-k_\perp^2$, respectively. And the normalization condition is
\begin{equation}
\int\frac{d\vec q}{(2\pi)^3}\frac{16\omega_1\omega_2\vec q^4}{15(m_1\omega_2+m_2\omega_1)}\left(-\frac{3\vec q^4}{M^2}f_3f_4-3\vec q^2f_3f_6 + 3\vec q^2f_4f_5+7M^2f_5f_6\right) = 7M.
\end{equation}

\end{document}